\documentclass{article}

\usepackage[nonatbib, preprint]{neurips_2020}

\usepackage[utf8]{inputenc} 
\usepackage[T1]{fontenc}    
\usepackage{hyperref}       
\usepackage{url}            
\usepackage{booktabs}       
\usepackage{amsfonts}       
\usepackage{nicefrac}       
\usepackage{microtype}      
\usepackage{graphicx}      
\usepackage{subfig}
\usepackage{natbib}
\usepackage{amsmath}
\usepackage{todonotes}

\title{Open Loop In Natura Economic Planning}

\author{%
  Spyridon Samothrakis \\
  Institute for Analytics and Data Science\\
  University of Essex\\
  Wivenhoe Park, Colchester CO4 3SQ \\
  \texttt{ssamot@essex.ac.uk} \\
}

\begin{document}

\newcommand{\io}{\texttt{IO-coeffs} }
\newcommand{\vpp}{\texttt{Vorpal Pick +1} }
\newcommand{\lc}{\texttt{Lucloelium} }
\newcommand{\tr}{\texttt{T-ring} }
\newcommand{\ol}{OLIN-EP}

\maketitle

\begin{abstract}
  The debate between the optimal way of allocating societal surplus (i.e. products and services) has been raging, in one form or another,  practically forever; following the collapse of the Soviet Union in 1991, the market became the only legitimate form of organisation --- there was no other alternative. Working within the tradition of Marx, Leontief, Kantorovich, Beer and Cockshott, we propose what we deem an automated planning system that aims to operate on unit level (e.g., factories and citizens), rather than on aggregate demand and sectors. We explain why it is both a viable and desirable alternative to current market conditions and position our solution within current societal structures. Our experiments show that it would be trivial to plan for up to 50K industrial goods and 5K final goods in commodity hardware. 
\end{abstract}

\section{Introduction}
With the advent of industrial revolution and the almost homicidal conditions that this ensued, ensued, voices arose which insisted there must be superior ways of organising society to the markets. The end and subsequent dissolution of the Soviet Union in 1991 killed most (if not all) ``living'' attempts to create a non-market economy. The rule of experts and democracy, the two major facets of modernism~\citep{scott1998seeing}, were to be permitted expression only through market mechanisms. As this paper is written, very few domains of social life remain outside the whims of the market and a state that serves it. Consequently, the totality of human experience (with obvious exceptions) is now part of market transactions. The victory of the market is so absolute that certain authors complain in the popular imagination: ``it is easier to envision the end of the world than the end of capitalism''~\citep{fisher2009capitalist}~\footnote{Some readers might complain that we conflate markets with capitalism. We are doing this on purpose.}. 

Within such a political and ideological onslaught, it is no surprise that research in alternatives (or partial alternatives) to the market remained very limited in scope. In this paper, we revisit one such alternative paradigm of societal distribution, whose invention (or inspiration) goes back quite some time ~\citep{marxII, clark1984planning, moseley1998marx}. We will provide a base for removing certain products from market circulation and provision them directly to citizens. The calculation of using products and services directly is generally called ``planning in natura''~\citep{cockshott2008calculation}, and has direct links to Universal Basic Services. The goal of planning methods is to remove the anarchy (and uncertainty) of production and provide citizens with consumption guarantees. Contrary to most of the authors we cite, our ambitions are somewhat social-democratic. We do not aim to replace the market, but instead focus on removing human reproduction from strictly ideological mechanisms. In fact, a conservative government not ``tied'' to market ideology could easily start implementing such a programme\footnote{We are not really that naive. Power politics abide.}. The goal of our specific programme is to match citizens and production units directly while monitoring the plan as closely as possible --- in order to take corrective action --- on a daily basis. Plan goals are to be formed using data collected from production units and citizens. 

We are not aware of any methods that attempt to plan production on the individual level, nor has there ever been an automated way to monitor the plan or amend it using data. The closest a quasi-automated system of planning that reached an (partial) operational level was Project Cybersyn \citep{beer1979heart}, but this was dismantled in a hurry following Pinochet's coup. Within the Soviet Union there is evidence that planning from final demand  was seen as a ``bourgeois'' \citep{bollard2019economists} and was never allowed, leaving production planning to the level of industrial goods (e.g., steel). The insistence to create plans and the focus of soviet economy to ``build machines that build machines'' might have contributed to the grim life of the soviet citizens in terms of consumer products. Prior to the late 1970s, when the demise of USSR became evident, some form of planning was always accepted within capitalist societies~\citep{judt2006postwar}. Japanese economists were effectively trained in planning by explicitly going through the works of Marx~\citep{karatani2020marx} until the late 80s. Our proposal of removing elements of production from market circulation is not historically controversial, but might look absurdly rebellious in a post-soviet world.

The rest of the paper is organised as follows; in Section \ref{background} we provide a generic discussion on the background and debate between economic planning and market economics, but also nudge at the link between economic planning, reinforcement learning and AI planning. Section~\ref{NEP} introduces a new model, which we term Open Loop In Natura Economic Planning. In Section~\ref{data} we discuss data collection issues --- and generally re-think the problem from the point of view of individual production units and citizens, while in Section~\ref{sec:sim} we perform a series of simulations. We conclude with a short discussion in Section~\ref{concl}.

\section{Planning vs the market}\label{background}


\subsection{Input-output economics and planning}
The problem of planning has been formally defined in \cite{lahiri1976input}, but we will attempt an RL based modernisation. Per unit of time $t$, a set of demands $d$ for certain goods (e.g, products, services) are to be satisfied for $c$ citizens. The planner's goal is to satisfy the demand of each citizen. In AI terms, we have something akin to a Markov Decision Process (MDP), with an agent (the planner) receiving information (the state) on the plan and a set of rewards related as to how closely the demand is met. Thinking of the problem as a single-player game-like MDP allows us to draw insights from the relevant literature, but it hides its complexity. The action space a planner would have to search through is massive - for $10^5$ individuals and $10^3$ goods, with each good having $10$ different quality levels for each individual, the planner would have to choose among $10^9$ real-valued actions (that exist on a very abstract level). Direct (AI) planning for this problem has never been considered, rather the effort has concentrated on strategic plans that operate on aggregate demand and sector level, with recalculations of the plan never taking place --- or, at best, on a yearly basis. 

The parent of modern mechanisms for planning (in this context) is what is termed the input-output model, which is thoroughly reviewed by \citet{leontief1986input}. The model comprises of an $nxn$ Matrix $A$ of technical coefficients, a vector $x$ of production level (i.e. how much we should produce for each product) and a demand vector $d$. The columns of the coefficient matrix conceptually ask the question ``how many units of each good to produce a single good of the type portrayed in this column do we need?''. The dot product of each row with the technical coefficients represents the consumption of a specific good. The demand vector $d$ represents how much external demand there is, i.e. that Equation~\ref{eq:basic} holds:

\begin{equation}
\centering
x_{i}=a_{{i1}}x_{1}+a_{{i2}}x_{2}+\ldots +a_{{in}}x_{n}+d_{i}
\label{eq:basic}
\end{equation}
    
In matrix notation, we have Equation~\ref{eq:matrix_basic}:

\begin{equation}
\centering
x = Ax + d \implies (I - A)x = d
\label{eq:matrix_basic}
\end{equation}

Something to note here is that traditional input-output models have no notion of time - all production is taking place within the same temporal unit. This is somewhat counterintuitive (and problematic for actual planning), but it allows a first easy approximation. It is the model proposed by \citet{cockshott1993towards}, covered by \citet{dyer2013red} and, with further additions (based on linear optimisation) discussed in ~\citet{cockshott2008calculation}. With no time element, the model remains suitable for very high level strategic planning - and indeed such models are widely used currently (e.g. most states publish input-output tables using monetary prices).

\subsection{Why planning?}

Von Mises and Hayek~\citep{von1935collectivist}, writing in the height of socialist revolutions, started putting together a critique of socialism, and more specifically (economic) planning. Parts of their critique (and this of their successors) sound still valid - for example same of their points on Marx's treatment of skilled vs unskilled labour\footnote{Marx's comment is that there are only quantitative differences in skilled vs unskilled labour}. Here we will concentrate on the arguments of planning using products and services (i.e. 10 kilos of rice, 20 pounds of flesh, 10 hours of electric supply) vs a market price allocation mechanism. Whether an optimal (automated or not) planner of such type could even exist is termed the \emph{calculation debate}. Arguments against the existence of an optimal planning mechanism fall into different camps, with some being aligned to moral questions (``it is unfair to just allocate goods'' or ``it is undemocratic''),  computational (``you can't compute the intermediate goods to produce'') or epistemic (``there is no way for the planner to know what to produce''). We will not discuss the democratic issue in this paper, though we strongly feel that the market is exceptionally undemocratic. It is now accepted by even the opponents of planning that computation should not be an issue ~\citep{brewster2004towards}. The epistemic argument, which is still very valid, entails that an optimal planner would not know \emph{what} to compute. A price mechanism would allow whoever is engaged with the market to express their preferences of goods in terms of how much they would be willing to pay, i.e. a very subjective preference function. Prices that (for producers) might, for example,  depend on the availability of goods~\citep{steele2013marx}.  In its extreme this holds true for consumers, as we have seen examples of iPad-for-kidney selling~\citep{student}, though we think it is safe to class such behaviours as pathological. If one makes the assumption of truly subjective values that vary continuously and are also widely different from person to person, then indeed a market might be able to allocate surpluses somewhat better than a plan. However, if you do accept that the majority of the population shares some similar preference function, at least in their top priorities (e.g. food, shelter, basic communication devices, electricity, health), the argument is nonsensical and applies only to incorporeal beings. Insofar as there are relatively slow changing patterns in consumption, standard machine learning models, combined with one's own predictions can be used to forecast demand.

\section{Open Loop In Natura Economic Planning}~\label{NEP}

Our method (Open Loop In Natura Economic Planning - \ol) builds upon the basic input-output framework. It creates a fundamentally different planning landscape than IO tables and is heavily inspired by current game playing / RL agents. The planning ``tick'' is no longer a year, but a day, and we expect the plan to be re-calculated based on observations and predictions every night. We no longer operate on abstract notions of aggregate demand, but instead we expect every individual to communicate their demands and projected demands daily. We also expect the productive units to recalculate their input-output coefficients (which we will call \io --- the values of the matrix $A$)  and provide them for plan updates on a daily basis in the form of a function --- more on this later. Closing, we maintain a notion of state that is missing from all original formulations. More formally, we operate on an MDP~\citep{puterman2014markov} that has the following characteristics:  

\begin{itemize}
    \item Actions $x \in \mathcal{A}$ capture what the production output of each industry should be. Note that due to notation conflicts with input-output literature we use $x$ for individual actions, rather than the most customary $a$.
    \item States $s \in S$ capture sufficient statistics of what we want to operate on, as transmitted every morning by production units and citizens. In our case, $s$ is simply a goods inventory.
    \item The transition function $T(s'|s,a)$ is formally unknown to us, but it is captured partially by the input-output matrix, partially by the semantics we give to the behaviour of different outputs of the matrix, and it operates on the inventory and externalities.
    \item The reward function denotes how happy the planner is in a specific state and is generally encoded as $R(s, a)$. We define later on a specific reward function that captures how well the plan targets are met and what damage the plan causes to the world.
    \item There is a discount factor $\gamma$, which attenuates closer vs further rewards. 
\end{itemize}

One can obviously claim that economic planning is more akin to a partially observable MDP (i.e. a POMDP), and this might be true, but unless one is to have the functions that describe the uncertainty over states, there is no reason to do the modelling this way. We could also start acting on histories of states and include externalities and rewards~\citep{izadi2005using}, but this might prove computationally infeasible. Claims could also be made that there is strong multi-agent element for the planner --- here we assume that everyone involved in the plan has it in their best interest to cooperate.

\subsection{The model}

We adapt a number of innovations to the standard input-output models, by changing the way we position the plan within the economy. As discussed before, the goal of an input-output matrix is to plan for demand at the end of a time period. Given that our goal is to provide necessities to sustain humans, we set all ``external'' demand to zero, and introduce a set of profiles combined with the number of citizens attached to each profile. You can see an example in Table~\ref{tab:planning}. Our input-output matrix describes the interactions between consumption profiles, a set of industrial goods,  and a set of final goods. Profiles are columns that describe the allocation of final goods to each citizen that has been assigned this specific profile.

\subsection{Nonlinearities and learning}

The plan formulation we described above inherits a number of limitations from the standard input-output model; the first one we will build upon is model linearity. The default model linearity is tremendously problematic --- for example there is the implicit assumption which is that labour needs will scale linearly with production demands. To address these issues, a generalisation of the input-output model~\citep{lahiri1976input,fujimoto1986non} looks as in Equation~\ref{eq:io}: 
\begin{equation}
\centering
 (I - F(x))x = d   
 \label{eq:io}
\end{equation}

This is profoundly liberating as a proposition, as we can stack production units and have different \io values as production scales. We can also extract from individual citizens how important hitting certain targets in their profile is.  Solving for $x$ now becomes a bit harder, as $F(x)$ could potentially be any function, but in our case, we constrain it to a specific matrix.  Remember that individual columns in the IO matrix represent how much it takes to produce a single unit of output --- it makes sense to define the matrix as in Equation \ref{eq:matrix} \begin{equation}
F(x) =\begin{bmatrix}{f_{00}}{\left(x_{0} \right)} & {f_{01}}{\left(x_{0} \right)}  & \dots &   {f_{0n}}{\left(x_{0} \right)} \\
{f_{10}}{\left(x_{1} \right)} & {f_{11}}{\left(x_{1} \right)}  & \dots &   {f_{1n}}{\left(x_{1} \right)} \\ 
\vdots & \vdots & \ddots & \vdots\\

\\{f_{n0}}{\left(x_{n} \right)} & {f_{n1}}{\left(x_{n} \right)} & \dots &  {f_{nn}}{\left(x_{n} \right)}
\label{eq:matrix}
\end{bmatrix}
\end{equation}

Constraining our function to this form has one important benefit; we can ask production units directly how many other goods they need in order to produce certain output units, and data scientists in these facilities can use any machine learning method to ``fit'' a curve and provide back a function..  

When it comes to the actual solution, one can attempt to use the gradient directly. The mean squared error $MSE((I-F(x))x, d)$ has a gradient that is $\nabla MSE((I-F(x))x, d) = 1/n\left((I-F(x))x - d\right)(I-F(x) - F'(x)x)$, which means that we can solve using any non-linear least squares algorithm --- or in fact any other non-linear optimisation algorithm. Another method (that comes from \citet{lahiri1976input}) is to go through the power series expansion $\left(I-A\right)^{-1}= \sum_{i=0}^\infty A^i  = I + A + A^2 + ...$ . We can then define $x_{(i+1)} = F(x_{(i)})x_{(i)} + d, x_{(0)} = d$ --- a recursive form of calculating $x$. This is what we are going to use in this paper, as it is based purely on linear solvers, and will find the global maximum as long as convexity is maintained. We could also attempt an end-to-end neural network solution (it is very easy to envision), but there are no (clear) advantages, unless a need arises to model exceptionally complex \io while optimising production at the same time, something we are not doing in this paper.

\subsection{Time and the transition function}
When it comes to producing goods and services, a model without a time element is severely limited; real production and consumption obviously have a time dimension. In the case of production, this is expressed in various forms like gestation times, production times, business inventories and depletion of resources. Multiple input-output models that include a time element have been developed\footnote{An example of such an Equation, from \citet{raa1986dynamic} is $x(t) = \sum_0^n\left[A_{t+s}(-s)x(t+s)\right]  + \sum_0^n\left\{B_{t+s+1}(-s)\left[x(t+s) + x(t+s + 1)\right]\right\} + z(t)$, with $A$ matrices representing circulating capital,  all $B$ matrices representing fixed capital, $z(t)$ is the demand at each point in time, while $-s$ is the ticks before the time $t$. }  \citep{raa1986dynamic,dobos2013dynamic,aulin1990dynamic} --- for an overview, see \citet{aulin2000dynamic},  The problem with these models is they were (for the most part) not designed with planning (in the AI sense) in mind. What we need to introduce (as discussed before) is a transition function $T(s'|s,a)$ and a notion of state $s$. This can really be anything that makes sense based on the individual components of what we have, but to simplify things we can define state as an inventory indicating how much we hold of everything we have so far, including any unwanted side effects (i.e. externalities) our methods are generating. The transition function now operates on that inventory/externalities vector, by adding things, removing things, showing when something is ready for consumption, and how much needs to be taken to gestation periods.

\begin{table}[]
    \centering
\begin{tabular}{lrrrrrr}
\toprule
{}                 Type &  Lucloelium &  Vorpal Pick +1 &  T-ring &  Profile 0 &  Profile 1 &  Demand \\
\midrule
         Lucloelium &       0.001 &           $f_{01}(x_0)$ &   1.000 &        3.0 &        2.0 &       0 \\
      Vorpal Pick +1 &       0.500 &           $f_{11}(x_1)$ &   0.000 &        0.0 &        0.0 &       0 \\
              T-ring &       0.000 &           0.000 &   0.000 &        0.1 &        0.2 &       0 \\
      Lb(Lucloelium) &       0.001 &           0.000 &   0.000 &        0.0 &        0.0 &       0 \\
  Lb(Vorpal Pick +1) &       0.000 &           0.012 &   0.000 &        0.0 &        0.0 &       0 \\
          Lb(T-ring) &       0.000 &           0.000 &   0.001 &        0.0 &        0.0 &       0 \\
           Profile 0 &       0.000 &           0.000 &   0.000 &        0.0 &        0.0 &       800 \\
           Profile 1 &       0.000 &           0.000 &   0.000 &        0.0 &        0.0 &       500 \\
\bottomrule
\end{tabular}

    \caption{Our example input-output matrix, for a society of 1300 citizens. Two of the \io vary with production levels - as there are three production units (see Figure~\ref{fig:production}) --- the rest are constant. Labour columns are omitted, as all values are zero.  There is one industrial good \vpp and two final goods. Demand now just signifies the number of individuals in each profile.}
    \label{tab:planning}
\end{table}

\subsection{Plan Humanity and externalities}

The goal of the plan is to deliver a set of products and services (termed goods in our setup) in real life, so the real rewards can only be measured when the plan has been executed. During the planning phase, however, we should have a reasonable indication of what is the level of rewards we have achieved. Let $\hat{d_i}(a_{ij} = 0)$ be the demand for a final good for a certain profile set to zero,  with $i$ coming from final goods $C$, while $j$ coming from profile consumption $P$. When we removed a good from a profile, we generate a surplus. That surplus, divided by how much that profile was expected to get, we define as the humanity of the plan. More formally, in Equation \ref{eq:HU} we define humanity $\mathcal{HU}_p$ as

\begin{align} 
    \mathcal{HU}^t_p &= \min_{i \in C, j \in P }\left\{   \left(\hat{d_i}(a_{i,j} = 0)\right)/\left(a_{ij}d_{j}\right)\right\} 
    \label{eq:HU}
\end{align}

Every profile created puts certain requirements on the economy in terms of unwanted side effects, commonly referred to as externalities (e.g. carbon from milk and meat production). We model externalities at each point in time as $\rho(e(x_t)x_t)$, with the total externalities for a plan being ${E}_p$ --- the sum of all externalities in time as in Equation~\ref{eq:ext0}, and $\rho$ being a function that weights the importance of each externality for each good: 

\begin{align} 
    \mathcal{E}_p^t &= \sum_0^t{\rho(e(x_t)x_t)} \label{eq:ext0} 
\end{align}

The difference between the way we measure the unwanted side-effects we get versus the goals we achieve is by design. In terms of production goals, a plan is as good as its worst performance. In terms of damage, we are measuring the cumulative effect --- we call this the Marc Anthony principle\footnote{``The evil that men do lives after them; The good is oft interred with their bones''}. A combination of externalities is what underlies the reward function.

\subsection{Plan execution}
Given that we do not have access to the real transition function (akin to training for a robot in an largely imperfect simulation), we suffer from two problems; first, that our plans are as limited in their ability to use future states as the imagination of the model creators. We will try to achieve certain goals every day for a year by following a set of actions that correspond to increasing production, without reference to future states - this is known as open loop planning - and is basically a vector $x$ per day. The fact that we re-plan on a daily basis means that we execute the plan in a closed loop setting - so overall we do \emph{open loop planning, closed loop execution}~\citep{bubeck2010open,weinstein2012bandit}. This is highly reminiscent of methods like Monte Carlo Tree Search~\citep{browne2012survey} that have shown tremendous success in games. The second problem is that the artificial conditions we optimise on might not correspond to reality. Again, this is a common problem in robotics and it is currently attacked by assuming fictional model hyperparameters, as to make the model robust~\citep{akkaya2019solving}.

\section{Data collection}\label{data}

The real world execution of the plan entails two steps: (a) The planner provides information to the production units on their daily targets and requests information on the previous day history, including \io in functional form and externalities. (b) The planner requests information on previous days demand and future demand from each individual (or discovers it).

\begin{figure*}[t!]
  \centering
  \includegraphics[keepaspectratio, width=\textwidth]{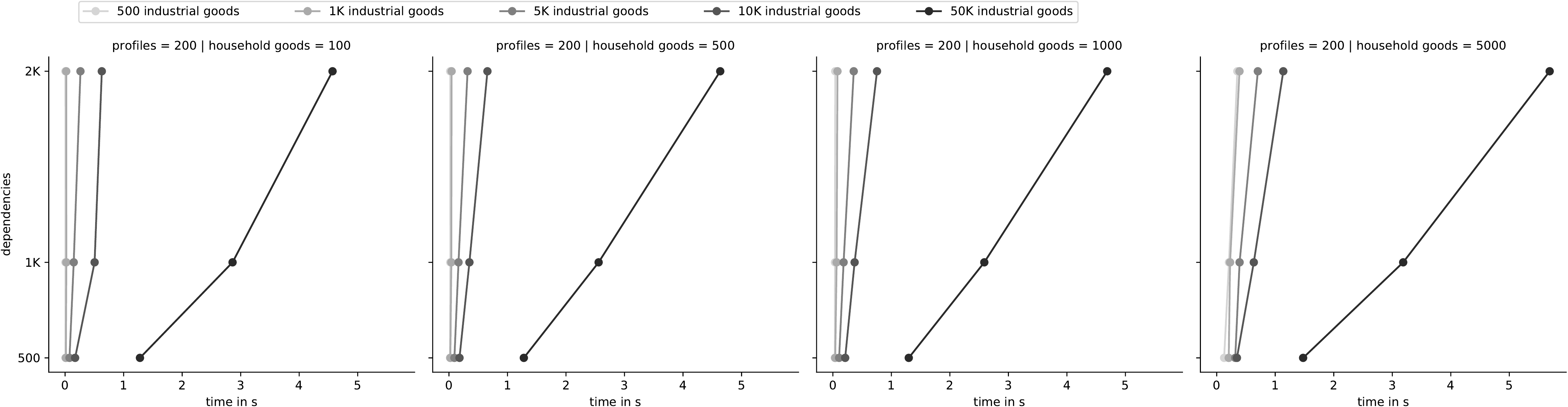}
  \caption{An example of performance scaling in solving the basic Equation of our model $(I-A)x = d$. Note that though $A$ is sparse, this does not follow that $x$ would be.}
  \label{fig:time}
\end{figure*}

\subsection{Production units}
Each production unit would have to effectively fill the columns of Matrix $F$ by providing the function $f_{ij}(x_i)x_{i}$, This can be achieved trivially by some form of active learning (i.e. asking managers: ``how much milk do you need to make one pound of cheese? How about two pounds? How about three?'') and interpolating accordingly. Alternatively, one can seed a classic ML model using past production data and combine it with active learning in any gaps. Now, converting these values into $f_{ij}(x_i)$ simply required dividing over the number of actual products $x_{i}$ for all possible values of $x_{i}$ . We expect production units to innovate constantly, achieving lower externalities and better \io, in a very organic process that amounts to optimisation coming from every part of the system.

\subsubsection{Citizens}
We have defined various profiles, but where do those profiles come from? This is essential --- these profiles are our reward function. Learning a reward function from consumption targets can be done by using any form of inverse reinforcement learning/preference learning on existing buying habits, direct questions and/or voting all in accordance with productive capacities. This should allow for effectively the discovery of basic needs on a fundamental level and the provision of relevant goods. From the outset, different profiles aim at addressing the problem of \textit{Variety}~\citep{beer1993designing} directly, i.e. we need to be able to act upon as many world states are possible. Individual profiles for every person would put tremendous strain on the planning mechanism and make the whole system very brittle, as any errors in production will result in a series of complaints. Instead, the focus should be on goods that allow for a high degree of customisation. For example, pre-packaged foods are a really bad production option, as they allow for very little tinkering. Allowing for very high degree of customisation and personalisation (i.g. a combination of (generative?) recipes plus food) should help make production both more robust and interesting. New types of computing devices, whose aim is to help so as to have the goods delivered be used in the most efficient and creative fashion possible, will also prove pivotal.

\subsection{Interactions with the market}
Given that the plan's aim is to complement, rather than abolish, the market, it is worth discussing what areas of production the plan will not shape. Goods in scarcity or products whose only value is their scarcity cannot be delivered through the plan; the subjectivity of the reward function would make it exceptionally hard to calculate individual preferences (and hence profiles), and would also open up the possibility of abuses, requiring constant vigilance to stop the creation of black markets. Goods in scarcity also open questions of multi-objective optimisation~\citep{erickson2013reason} --- that will mostly lead to a wealth of equally non-satisfactory solutions. Any invention that helps the plan should be readily adapted. New products and services could also come from market forces. This would require the market to turn into activities that look more like prospecting --- anything that a plan cannot cover should generate profit. The most important point, however, when it comes to market, is not to allow it to use the plan as a way of undercutting wages; once the plan is introduced, it should be followed by a policy of \textit{increasing minimum wages and decreasing working hours, in accordance with productivity gains} in order to start removing human labour from the market and reaping benefits from further automation. For example, shoe production is still a very manual process, and high wages in the sector should come from automation. 

  

\begin{figure*}[!tbp]
  \centering
  \subfloat[An example of how much \lc and \vpp is needed to create units \vpp as portrayed in the x axis - i.e.  $f_{ij}(x_i)x_{i}$.]{\includegraphics[trim={0cm 0cm 0 0cm}, width=0.5\textwidth]{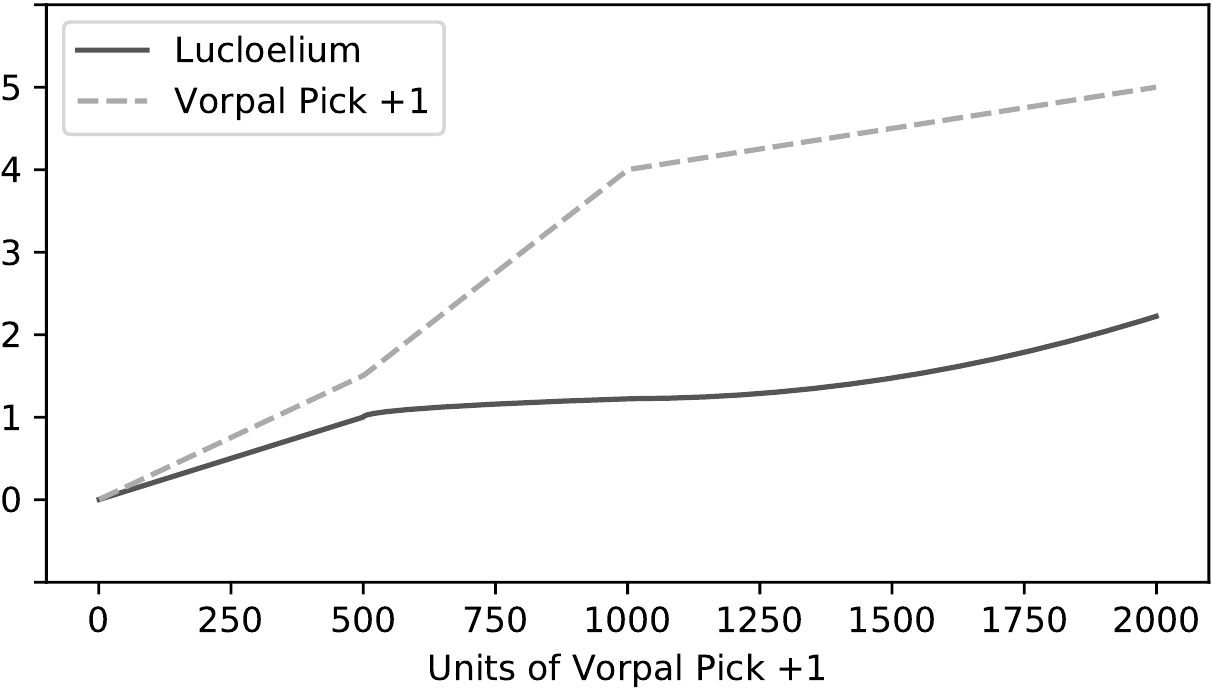}\label{fig:f1d}}\subfloat[ $f_{ij}(x_i)$ for \lc and \vpp.]{\includegraphics[trim={0cm 0cm 0 0cm}, width=0.5\textwidth]{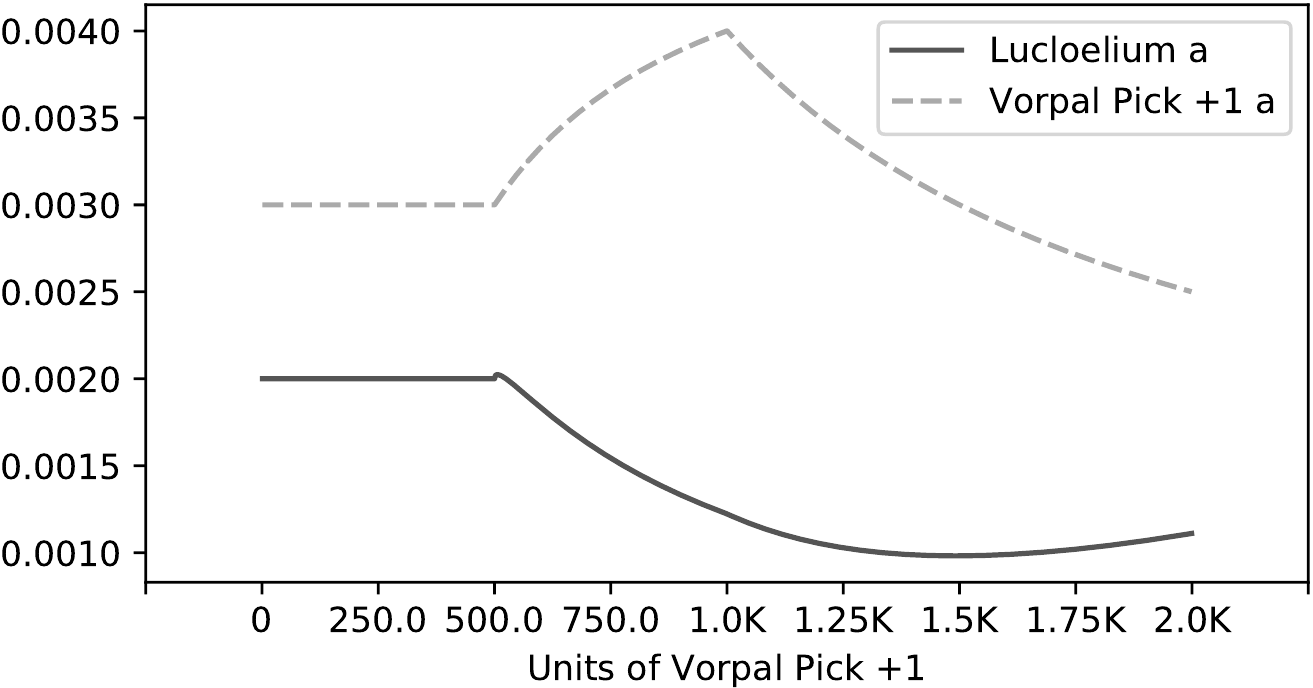}\label{fig:f2d}}
  \caption{$f_{01}(x_0)$ and $f_{11}(x_1)$ derivation from production outputs. There are three fictional production units that follow very different curves in their models.}
  \label{fig:production}
\end{figure*}

\section{Simulations}\label{sec:sim}

\begin{figure*}[t]
  \centering
  \subfloat[A simulation without noise ]{\includegraphics[trim={0cm 0cm 0 0cm}, width=0.5\textwidth]{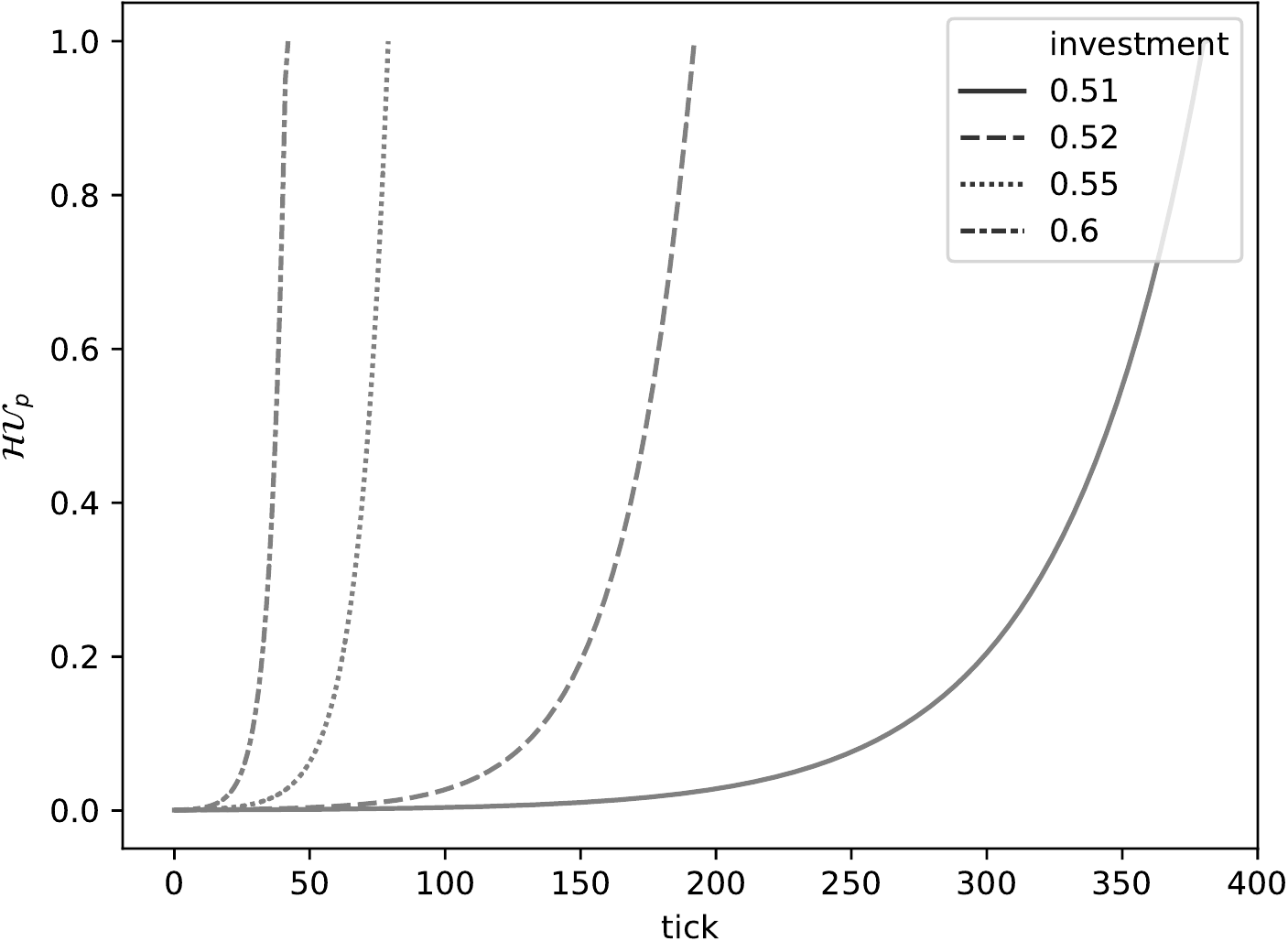}\label{fig:f1}}\subfloat[A simulation with noise - certain investment profiles fail to achieve self-sustainability.]{\includegraphics[trim={0cm 0cm 0 0cm}, width=0.5\textwidth]{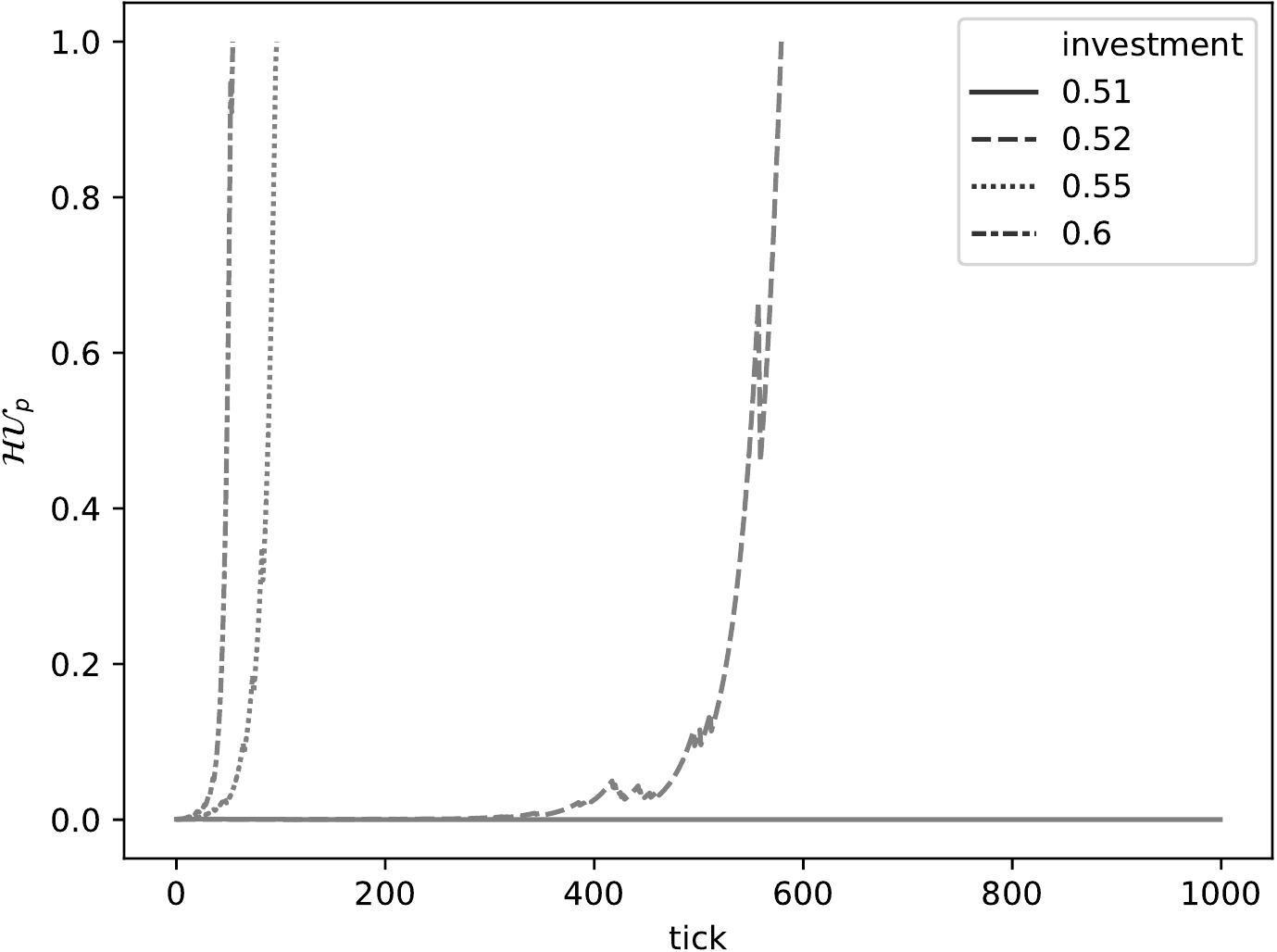}\label{fig:f2}}
  \caption{Humanity of the plan vs investment profiles. Note the exponential curves. }
 \label{fig:simulations}
\end{figure*}

We performed a number of simulations on imaginary data. The first set of simulations resolves around solving $(I-A)x = d$ repeatedly for matrices of different size. Solving this set of linear Equations fast is fundamental as both our time element and the non-linearity solution depend it. We have run all possible combinations of industrial goods (i.e. goods not needed by the profiles, $ [500,1000, 5000, 10000, 50000 ]$, final goods of $[50, 100, 500,1000, 5000]$, a profile of size $200$ (i.e. 200 different combinations of final goods),  with each good needing  $[500,1000, 2000]$ other goods in order to be made. The results can be seen in Figure \ref{fig:time} --- all results were collected on a \texttt{CPU: Intel i7-8700K @ 4.800GHz / 64GB RAM}, using scipy~\citep{2020SciPy-NMeth}. Alternative solutions that include gradient estimations might be faster, but this will probably depend on the problem. As it stands, the deciding speed factor is the number of dependencies, but everything is solved in well below 20 seconds. Overall, it is trivial to attack the problem, 

We also simulate a sample, completely fictional economy of an alien village, in Table \ref{tab:planning}. The economy is made up from two final goods (\lc,\tr) and one industrial good (\vpp). The initial quantities of each item in inventory are restricted. The results of the simulation can be seen in Figure \ref{fig:simulations}. The village plans to provide the final goods in two profiles. The village starts without being able to fulfil the goals of each profile, hence they are forced to produce a limited amount of goods at each daily tick and invest the rest. What this means in practical terms is that units of \lc we create get ``consumed'', while units of \vpp just get added on. Notice the exponential rise in humanity of the plan. We perform a second experiment, where with a certain probability a portion of the inventory would just vanish. Here (see Figure \ref{fig:simulations}(b)) lower investment leads to collapse, with the humanity of the plan never recovering. This effect would not be visible without including some noise to the model. Finally, also note that the only real difference between a simulation and a plan comes from the fact that we think that the simulation is closer to reality --- there is no way to execute it in real life.

\section{Conclusion}\label{concl}
Lenin is quoted and attacked directly by Von Mises in \citet{von1935collectivist}. His remarks concern the efficacy of the Leninist take on society --- planning is more or less an ideologically motivated position put forward by a dictator. We think the Tables might have turned --- if anything, with the current development of the means of production the market looks more and more like an ideological tool and a vampiric dictator, lurching from crisis after crisis. At the same time,  a ``no-future'' rhetoric seems to be becoming the norm. The market (and in turn, capital) operate on humans indirectly, the market ``reward'' function does not have to do with meeting needs, but rather with profit maximisation. Insomuch, it makes human lives an externality. Our method, \ol, simply combines a set of production units with citizens' basic needs in an RL-like format. Putting together a full planning programme for basic goods is not trivial, but it is painfully obvious that the technological tools have been there for some time. We hope that this paper re-starts the discussion on a technical level, with ever increasing planning methods and simulations coming to light. There is no reason for the plan to be as simple as the one discussed here --- in fact Facebook is currently performing large scale simulations \citep{ahlgren2020wes}; one can easily  envision a situation where simulated production/consumption unit behaviour is used to plan in a setting more closely aligned to traditional RL.

\cleardoublepage

\section{Broader Impact}
The whole of this article is essentially societal commentary, proposing a new production system for basic products and services. If what we propose is adapted, even if not in the exact form we discuss above, it will change the way society works forever. When authors deliberate on the impact of Artificial Intelligence or Machine Learning on society, they follow a line almost always bound within a liberal framework (e.g. how to make sure women and men have the same opportunities of getting ``good jobs'' ). We think it is high time to move beyond liberalism when discussing technological impact. 

\bibliographystyle{plainnat}
\bibliography{biblio.bib}

\end{document}